\documentclass[aps,twocolumn,showpacs]{revtex4}
\usepackage{graphicx,amsmath,amssymb}
\newcommand{\figwidth}{0.95\columnwidth}
\newcommand{\eq}[1]{Eq.(\ref{#1})}
\newcommand{\fig}[1]{Fig.~\ref{#1}}
\newcommand{\avg}[1]{ {\langle #1 \rangle} }
\newcommand{\olcite}[1]{Ref.~\onlinecite{#1}}
\newcommand{\ahum}[1]{``#1''}

\newcommand{\zcr}{z_{\rm cr}}

\newcommand{\beq}{\begin{equation}}
\newcommand{\eeq}{\end{equation}}
\newcommand{\bea}{\begin{eqnarray}}
\newcommand{\eea}{\end{eqnarray}}

\begin{document}

\title{The Widom-Rowlinson mixture on a sphere : Elimination of 
exponential slowing down at first-order phase transitions}

\author{T. Fischer and R. L. C. Vink}

\affiliation{Institute of Theoretical Physics, Georg-August-Universit\"at 
G\"ottingen, Friedrich-Hund-Platz~1, 37077 G\"ottingen, Germany}

\date{\today}

\begin{abstract} Computer simulations of first-order phase transitions using 
\ahum{standard} toroidal boundary conditions are generally hampered by 
exponential slowing down. This is partly due to interface formation, and partly 
due to shape transitions. The latter occur when droplets become large such that 
they self-interact through the periodic boundaries. On a spherical simulation 
topology, however, shape transitions are absent. By using an appropriate bias 
function, we expect that exponential slowing down can be largely eliminated. In 
this work, these ideas are applied to the two-dimensional Widom-Rowlinson 
mixture confined to the surface of a sphere. Indeed, on the sphere, we find that 
the number of Monte Carlo steps needed to sample a first-order phase transition 
does not increase exponentially with system size, but rather as a power law 
$\tau \propto V^\alpha$, with $\alpha \approx 2.5$, and $V$ the system area. 
This is remarkably close to a random walk for which $\alpha_{\rm RW}=2$. The 
benefit of this improved scaling behavior for biased sampling methods, such as 
the Wang-Landau algorithm, is investigated in detail.\end{abstract}


\pacs{05.70.Fh, 82.20.Wt, 05.70.Np}

\maketitle

\section{Introduction}

\begin{figure}
\begin{center}
\includegraphics[width=\figwidth]{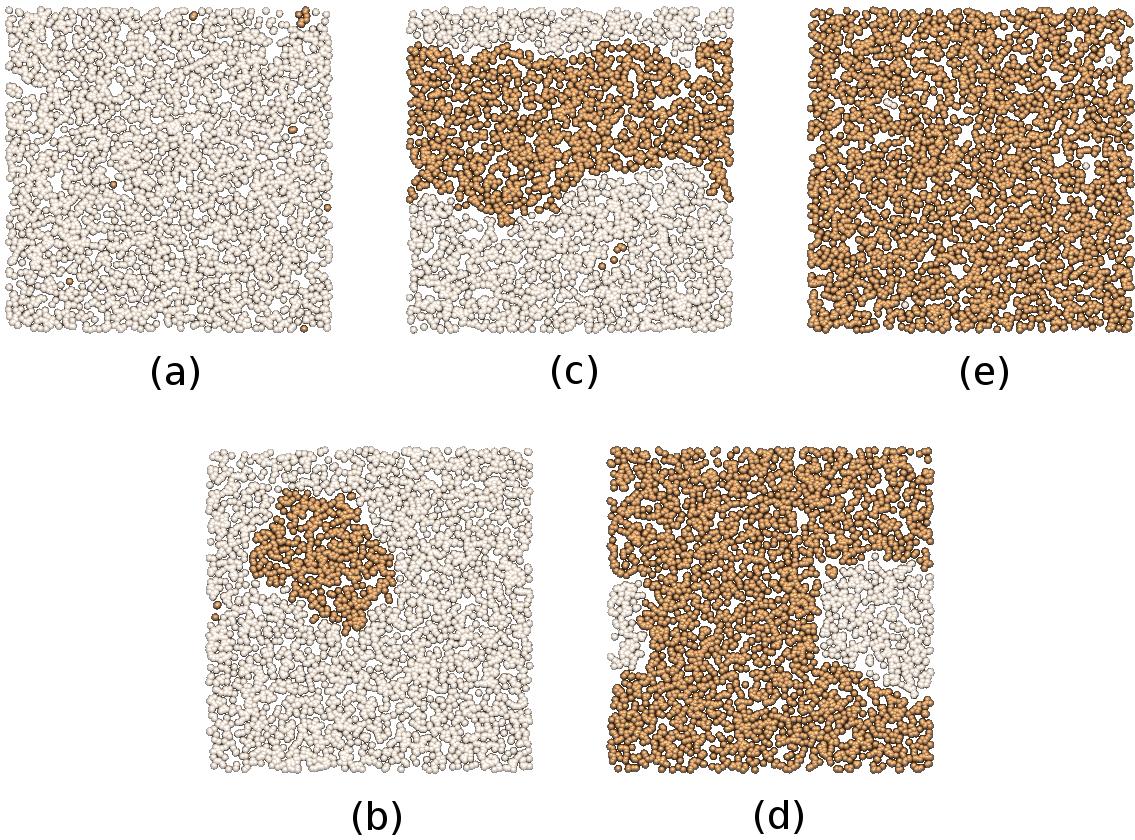}
\caption{ 
Phase separation snapshots of the 2D WR mixture as obtained in grand canonical 
MC simulations using \ahum{standard} toroidal boundary conditions, i.e.~a 
square with periodic boundaries. The snapshots were obtained at fugacity 
$z=2.5$, which is well above the critical fugacity $\zcr$, and so the transition 
is strongly first-order here. The light regions correspond to the vapor phase, 
dark regions to the liquid phase. Starting in the vapor phase (a), a droplet of 
liquid nucleates (b). The droplet grows until the strip configuration is reached 
(c). Further increasing the amount of liquid phase leads to a droplet of vapor 
(d), and finally this droplet vanishes, leading to the pure liquid phase (e).}
\label{fig:torus_sysview}
\end{center}
\end{figure}

Phase transitions in colloidal suspensions are of profound practical interest. 
Think, for instance, of phase separation in colloid-polymer mixtures 
\cite{aarts.schmidt.ea:2004}, or the freezing of colloidal hard spheres at high 
densities \cite{poon:2004}. For this reason, there is also an enormous interest 
in the modeling of phase transitions by means of computer simulation. The 
investigation of phase transitions via computer simulation is not trivial, as 
there are numerous hurdles to overcome. One obvious problem is the issue of 
finite system size. Since computational resources are limited, one always deals 
with finite numbers of particles, whereas phase transitions are defined in the 
thermodynamic limit. Hence, there is an obvious \ahum{gap} to bridge, achieved 
in practice using finite-size scaling \cite{fsslit}.

Another problem, which we focus on in the present paper, concerns exponential 
slowing down, and affects simulations of first-order phase transitions 
\cite{berg.neuhaus:1992}. To illustrate the problem we consider phase separation 
in the Widom-Rowlinson (WR) mixture \cite{widom.rowlinson:1970} in two 
dimensions (2D). In this model there are two particle species, A and B, each 
modeled as disks of diameter $a$ (in what follows $a$ will be the unit of 
length). The only interaction is a hard-core repulsion between A and B disks. As 
is well known, the WR mixture phase separates provided the fugacities $z_A$ and 
$z_B$, of A and B particles, are high enough; due to symmetry it holds that $z_A 
= z_B \equiv z$ at the transition, which we shall use throughout this work. When 
phase separation occurs, one obtains a \ahum{vapor} phase (dense in B species, 
and lean in A species) and a \ahum{liquid} phase (lean in B species, and dense 
in A species). The particle density of A species $\rho_A = N_A/V$ may be used as 
order parameter to distinguish between the phases, since it assumes a low value 
in the vapor phase, and a high value in the liquid. Here, $N_A$ is the number of 
A particles in the system, and $V$ the total system area (the choice for A or B 
is arbitrary, of course). Despite its simplicity, the 2D WR mixture is relevant 
for phase separation in cell membranes, as the latter also constitute 
effectively 2D systems (which even give rise to 2D Ising critical exponents 
\cite{citeulike:3850776}).

The standard approach to simulate the 2D WR mixture is to perform a grand 
canonical Monte Carlo (MC) simulation on a $V = L \times L$ square with 
periodic boundaries. Provided the fugacity $z$ significantly exceeds the 
critical value $\zcr$, such that the transition is first-order, phase separation 
proceeds as shown in \fig{fig:torus_sysview} \cite{citeulike:5076327, 
citeulike:4503186}. Starting in the vapor phase (a), a nucleation event occurs, 
leading to the condensation of a droplet of the liquid phase (b). The droplet 
grows until it interacts with itself through the periodic boundaries, leading to 
the strip configuration (c). In the strip configuration, vapor and liquid 
coexist with each other, separated by two interfaces that run perpendicular to 
one of the edges of the simulation square as this minimizes the interfacial 
area. The approach to the liquid proceeds via the formation of a vapor droplet 
(d), which eventually vanishes, leading to a pure liquid phase (e).

The path connecting vapor and liquid thus passes the strip configuration of 
\fig{fig:torus_sysview}(c). However, in a standard MC simulation, where 
configurations appear proportional to their Boltzmann weight, the strip 
configuration is extremely rare, due to the large amount of interface that it 
contains. In 2D, the total interface length in the strip configuration equals 
$2L$, corresponding to a free energy barrier $\Delta F = 2 \sigma L$, where 
$\sigma$ is the line tension. Hence, starting in one of the pure phase (a) or 
(e), it typically takes $\tau \sim \exp \left( 2 \sigma L \right)$ MC steps to 
reach the strip configuration. Since the \ahum{tunneling} time $\tau$ increases 
exponentially with the system size $L$, and hence explains the phrase 
\ahum{exponential slowing down}, it is clear that the standard MC method must be 
modified at a first-order phase transition in order to remain efficient.

Such modifications have been made, and the state-of-the-art is to not sample 
from the Boltzmann distribution, but from a modified distribution, such that the 
\ahum{unfavorable} interface configurations of \fig{fig:torus_sysview}(b,c,d) 
are sampled just as often as the \ahum{pure} phases (a) and (e). Crucial to 
these methods is the use of an order parameter, constructed such that
\begin{enumerate}
\item[(1)] it varies strictly monotonically along the path connecting the 
phases, and
\item[(2)] is computationally fast to calculate. 
\end{enumerate}
Regarding the 2D WR mixture, a convenient order parameter is the particle 
density $\rho_A$, which certainly fulfills criterion (2). In the vapor phase 
$\rho_A \equiv \rho_{\rm A,vap}$, in the liquid phase $\rho_A \equiv \rho_{\rm 
A,liq}$, while in the strip configuration $\rho_A \approx (\rho_{\rm A,vap} + 
\rho_{\rm A,liq})/2$. The idea is to perform a MC simulation using a biased 
energy $E_B = E_0 + w(\rho_A)$, with $E_0$ the original energy of the system, 
and $w(\rho_A)$ some {\it a priori} unknown function of the order parameter 
$\rho_A$. Clearly, by tuning $w(\rho_A)$ appropriately, the probability of the 
interface configurations can be artificially enhanced. The aim is to construct 
$w(\rho_A)$ such that the simulation performs a random walk in $\rho_A$. Various 
methods can be used to construct $w(\rho_A)$ in practice, such as multicanonical 
sampling \cite{berg.neuhaus:1992}, (successive) umbrella sampling 
\cite{citeulike:2414151, virnau.muller:2004}, and Wang-Landau sampling 
\cite{wang.landau:2001}. The methods differ in details, but all provide a means 
to obtain $w(\rho_A)$.

Hence, using a suitable $w(\rho_A)$, the free energy barrier of interface 
formation is eliminated, the coexistence configurations of 
\fig{fig:torus_sysview}(b,c,d) become accessible, and a random walk in $\rho_A$ 
will result (or so one hopes). In practice, this is not the case 
\cite{citeulike:4503186} because, strictly speaking, $\rho_A$ does not fulfill 
criterion (1) and hence is not a suitable order parameter. To see this, consider 
the case where half of the simulation square is occupied with vapor, and the 
other half with liquid. One way to arrange the phases is the strip configuration 
(c), with the distance between the interfaces being $L/2$. However, one could 
equally well arrange the phases in one of droplet configurations (b) or (d), 
with the droplet radius being $L / \sqrt{2 \pi}$. Both \ahum{solutions} yield 
the same order parameter $\rho_A$, but clearly differ in topology. This is a 
problem because the transition from the droplet to the strip configuration is a 
first-order transition by itself, with a complicated order parameter not simply 
related to $\rho_A$ \cite{citeulike:4503186}. As with any first-order 
transition, these so-called {\it shape transitions} also lead to exponential 
slowing down. This means that simulations do not yield random walk behavior in 
the order parameter, even when $w(\rho_A)$ is accurately known. Instead, most 
time is spend in the pure phases (a) and (e), or in the strip configuration (c), 
but transitions between the pure phases and the strip become increasingly rare 
with increasing system size \cite{citeulike:4503186, vink.schilling:2005}. This 
leads to poor sampling statistics in practice.

\begin{figure}
\begin{center}
\includegraphics[width=\figwidth]{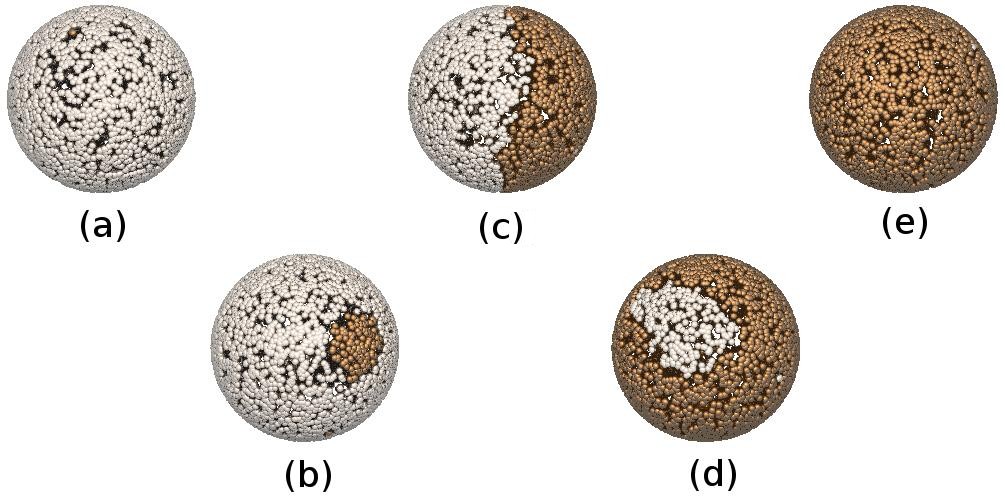}
\caption{ 
The analogue of \fig{fig:torus_sysview} but this time the simulation was 
performed on the surface of a sphere. In contrast to toroidal boundary 
conditions, the transition from the vapor (a) to the liquid (e) involves only 
the nucleation of a droplet. Once such a droplet has formed, the path from (b) 
to (d) does not involve any shape transitions; the images (b), (c), and (d) 
only differ in the relative volume of the phases, not in their geometric 
configuration.}
\label{fig:sphere_sysview}
\end{center}
\end{figure}

In principle, these problems can probably be overcome using a more sophisticated 
order parameter, one that also distinguishes between shape. However, such order 
parameters are not trivial to construct, and are likely to be computationally 
expensive, which would violate criterion (2). An alternative approach is to use 
a different simulation box topology, one in which shape transitions are absent 
\cite{citeulike:4503186, citeulike:5076327}. Note that the droplet-strip 
transition is a consequence of using a square simulation box with periodic 
boundaries (topologically equivalent to a torus). It is clear that on the 
surface of a sphere, the droplet-strip transition would not occur. Instead, on a 
sphere, we expect a first-order transition to proceed as shown in 
\fig{fig:sphere_sysview}. Starting in the vapor phase (a), we still expect a 
nucleation event, leading to a droplet of the liquid phase (b). By increasing 
$\rho_A$ further, the droplet grows continuously (c,d) but there are no shape 
transitions. There is still one nucleation event, of course, as the vapor 
droplet (d) vanishes, leading to a pure liquid phase (e). Hence, on the surface 
of a sphere, shape transitions are absent, and by using an accurate bias 
function $w(\rho_A)$ one should achieve behavior more closely resembling a 
random walk in $\rho_A$. In fact, any remaining slowing-down is due to 
nucleation, i.e.~of actual physics taking place, and should prove rewarding to 
study further.

The primary aim of this paper is to investigate if exponential slowing-down at 
first-order transitions is indeed eliminated on a spherical topology. The idea 
of doing so was announced some time ago \cite{citeulike:5076327}, but as far as 
we know, such simulations have not been performed to date. In fact, we are only 
aware of \olcite{citeulike:4503186}, which considers a 2D Ising model on the 
surface of a cube. The latter only crudely approximates a sphere, but already 
the barriers arising from shape transitions were seen to soften. Of course, 
being a lattice model, the extension to a spherical topology is not feasible in 
the Ising case. In contrast, for the 2D WR mixture, which is entirely {\it 
off-lattice}, the extension to a spherical topology poses no fundamental 
objections.

The outline of this paper is as follows. We first describe the grand canonical 
MC method in the presence of a bias function, and we provide some implementation 
details as to how this method can be efficiently implemented on the surface of a 
sphere. Next, we perform a number of cross-checks, to demonstrate that toroidal 
and spherical simulation topologies are consistent with each other in the 
thermodynamic limit. We then turn to our main result, and show that at 
first-order phase transitions, exponential slowing down on the sphere is largely 
eliminated. Finally, we discuss how this improved performance benefits a number 
of sampling algorithms, in particular the Wang-Landau algorithm. We end with some 
concluding remarks in the last Section.

\section{Methods}

\subsection{Grand canonical Monte Carlo}
\label{sec:bias}

We simulate the 2D WR mixture in the grand canonical (GC) ensemble using a bias 
function $w(N_A)$, i.e.~defined on the number of A particles. The WR mixture was 
defined in the Introduction; here we only explain the GC simulation method. In 
the GC ensemble, the fugacity $z$ and the system area $V$ are fixed, while the 
number of particles fluctuates. Phase separation is studied using the order 
parameter distribution $P_V(N_A|z)$, defined as the probability to observe a 
system containing $N_A$ particles of type A. We emphasize that $P_V(N_A|z)$ 
depends on the imposed fugacity $z$, the system area $V$, and on the simulation 
box topology (here: toroidal or spherical). The basic MC steps used to sample 
the distribution are insertions and removals of single particles. At each step, 
the simulation attempts with equal probability one of the four following moves:
\begin{enumerate}
 \item[(1)] insertion of one A particle at a random location,
 \item[(2)] removal of one randomly selected A particle,
 \item[(3)] insertion of one B particle at a random location,
 \item[(4)] removal of one randomly selected B particle.
\end{enumerate}
If the insertion attempts lead to forbidden overlaps, they are rejected. 
Otherwise, the moves are accepted with probabilities
\bea
 \label{eq:AccInsA}
 p_{\text{ins,A}} 
   &=& 
 \min \left[1, \frac{zV}{N_A+1} 
 e^{w(N_A+1)-w(N_A)} \right], \\
 \label{eq:AccRemA}
 p_{\text{rem,A}} 
   &=& 
 \min \left[1, \frac{N_A}{zV}
 e^{w(N_A-1) - w(N_A)} \right], \\
 p_{\text{ins,B}} 
   &=& 
 \min \left[1, \frac{zV}{N_B+1} \right], \\
 p_{\text{rem,B}} 
   &=& 
 \min \left[1, \frac{N_B}{zV} \right].
\eea
In the above, $N_A$ ($N_B$) refers to the number of A (B) particles in the 
system at the start of the move. Note the presence of the bias function $w(N_A)$ 
in moves involving $A$ particles. As was explained in the Introduction, the bias 
function is needed to overcome the free energy barrier of interface formation. 

\subsection{Implementation}

The most CPU consuming steps are particle insertions, since here one needs to 
check for overlap with particles of the opposite species. We now discuss how 
these checks can be performed efficiently on the surface of a sphere. To 
simulate a total area $V$, the sphere radius must be $R = \sqrt{V/4\pi}$. The 
position of each particle on the sphere is stored using a 3D vector $\vec{r} = 
(x,y,z)$ with $|\vec{r}|=R$. This means carrying around a third coordinate but 
allows to eliminate time-consuming trigonometric functions. First note that the 
on-sphere distance $d$ between two particles $i$ and $j$ is the length of the 
shortest path over the sphere. This path lies on a great-circle, i.e.~a circle 
with radius $R$. Hence, $d = R \, \theta$, where $\cos \theta = \left( \vec r_i 
\cdot \vec r_j \right) / R^2$. Unlike particles $i$ and $j$ overlap when 
$d<a$, with $a$ the particle diameter or, equivalently, whenever
\beq
 \vec{r}_i \cdot \vec{r}_j > R^2 \cos \left( a/R \right),
\eeq
where the right-hand side is a constant, which needs to be evaluated only once 
at the start of the simulation. Hence, to check for overlap, only the 
computationally cheap term $\vec{r}_i \cdot \vec{r}_j$ is needed but no 
trigonometric functions.

The second optimization concerns the implementation of link-cell neighbor lists 
\cite{allen.tildesley:1989} on the sphere. To this end, the sphere (of radius 
$R$) is \ahum{embedded} in a 3D cube of edge $2R$. The cube itself is 
partitioned in $n \times n \times n$ equally sized sub-cubes, with $n=2R/a$ 
rounded down to the nearest integer. Since it holds that 
\beq\label{ineq:distances}
 d \geq | \vec r_i - \vec r_j |,
\eeq
with $d$ the on-sphere distance between particles $i$ and $j$, one only needs to 
check for overlap with particles that are in the same sub-cube, or in any of the 
neighboring sub-cubes (including diagonal neighbors). Note that the number of 
neighboring sub-cubes that needs to be checked is typically less than the 
maximum of $3^3-1=26$ possible neighbors, since only sub-cubes actually 
intersecting with the surface of the sphere have to be taken into account. In 
practice, only about $13-14$ neighboring sub-cubes were counted in our 
simulations.

\section{Results}

\subsection{Order parameter distribution and line tension}

\begin{figure}
\begin{center}
\includegraphics[width=\figwidth, viewport=55 75 750 550]{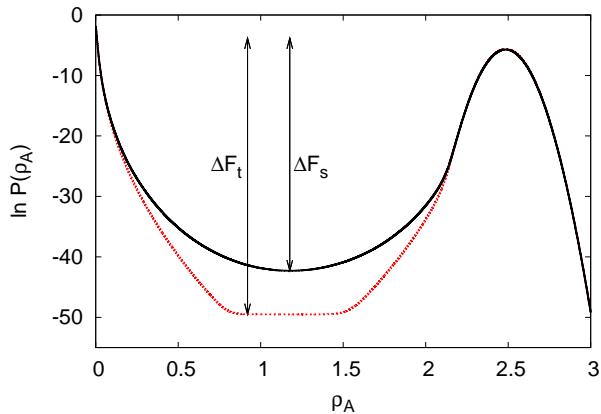} 
\caption{Order parameter distribution $P_V(N_A|z)$ for the 2D WR mixture 
obtained at fugacity $z=2.5$ using toroidal (dashed line) and spherical (solid 
line) simulation topologies. Note the logarithmic scale. The system area equals 
$V=1600$ in both cases. Due to the high value of $z$, the vapor peak on the left 
is squeezed to the edge. Also indicated are the free energy barriers $\Delta 
F_t$ and $\Delta F_s$, which can be used to obtain the line tension $\sigma$.}
\label{fig:probdist}
\end{center}
\end{figure}

\begin{figure}
\begin{center}
\includegraphics[width=\figwidth, viewport=55 75 750 550]{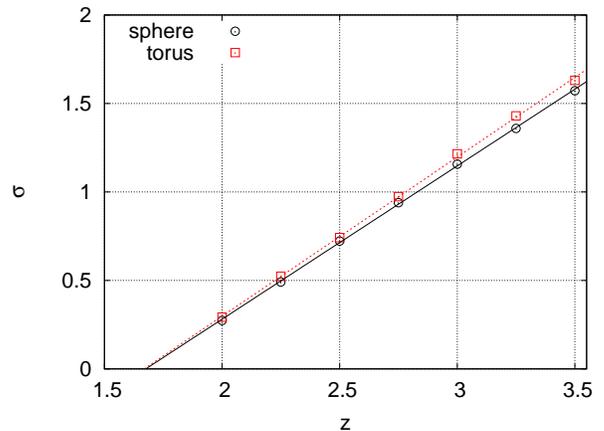} 
\caption{ Variation of the line tension $\sigma$ versus $z$ as obtained using 
toroidal (dashed line with squares) and spherical (solid line with circles) 
simulation topologies. The data were obtained at fixed system area $V=1600$ in 
both cases, and so finite-size effects are not accounted for.}
\label{fig:linetension}
\end{center}
\end{figure}

We begin our analysis by explicitly showing the order parameter distribution 
$P_V(N_A|z)$ as obtained using a toroidal and spherical simulation topology. 
\fig{fig:probdist} shows a typical result for a fugacity $z$ high above the 
critical fugacity $\zcr$; the system area $V$ is the same in both cases. The 
distributions reveal two peaks: the left peak corresponds to the pure vapor 
phase, the right peak to the pure liquid, and from the peak positions $\rho_{\rm 
A,vap}$ and $\rho_{\rm A,liq}$ can be read-off. Whenever the simulation visits 
the peaks, a single homogeneous phase is observed, i.e.~resembling the snapshots 
of \fig{fig:torus_sysview}(a,e) and \fig{fig:sphere_sysview}(a,e). On the scale 
of the graph, the peak positions between the toroidal and spherical topology 
practically coincide. This is to be expected because the pure phases do not 
contain any interfaces. Only in the region between the peaks are differences 
between the two topologies expected to appear.

For the toroidal topology, a pronounced flat region between the peaks unfolds. 
This is where the system assumes the strip configuration of 
\fig{fig:torus_sysview}(c). In this configuration, an increase of the volume of 
either phase at the expense of the other phase merely moves the interface but 
does not change its form nor affect the bulk phases. Hence, the free energy 
remains the same under such a change which is the origin of the characteristic 
flat region in the probability distributions of systems on a toroidal topology. 
Following Binder \cite{binder:1982}, the average height $\Delta F_t$ of the 
peaks above the flat region, measured in the logarithm of $P_V(N_A|z)$, 
corresponds to the free energy cost of interface formation. Since the total 
amount of interface in the strip configuration equals $2L$, one immediately 
obtains the line tension
\begin{equation}\label{eq:lt}
 \sigma = \Delta F_t / (2L) \hspace{5mm} \mbox{(toroidal topology)},
\end{equation}
with $L$ the edge of the simulation square. In contrast, using a spherical 
topology, $P_V(N_A|z)$ does not reveal any flat region. The reason is that on 
the sphere, any change in the relative volume of the phases inevitably creates 
or destroys interface. The maximum amount of interface is generated when half 
the sphere is occupied with vapor, and the other half with liquid, i.e.~conform 
\fig{fig:sphere_sysview}(c). The interface length then equals $2 \pi R$, and the 
analogue of \eq{eq:lt} becomes
\begin{equation}
 \sigma = \Delta F_s / (2 \pi R) \hspace{5mm} \mbox{(spherical topology)},
\end{equation}
with $R$ the sphere radius, and $\Delta F_s$ the barrier height. The estimates 
of the line tension as a function of the fugacity are shown in 
\fig{fig:linetension} for both topologies using $V=1600$. In agreement with 
theoretical expectations, $\sigma$ decreases with decreasing $z$, and at $\zcr$ 
it vanishes. However, $\sigma$ obtained on the sphere is systematically below 
that of the torus, the discrepancy being around 5\%. In principle, we only 
expect agreement in the thermodynamic limit, and so a detailed investigation of 
finite-size effects \cite{berg.hansmann.ea:1993} is required to resolve this 
issue. In addition, $\sigma$ obtained on the torus corresponds to planar 
interfaces, whereas the interfaces on the sphere are curved. Hence, there could 
be curvature corrections, possibly involving Tolman's length 
\cite{citeulike:4609411}.

\subsection{Locating the critical point}

\begin{figure}
\begin{center}
\includegraphics[width=\figwidth, viewport=55 75 750 550]{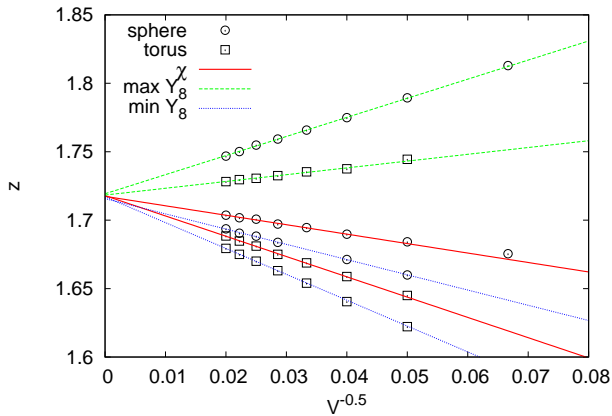}
\caption{Determination of $\zcr$ via finite-size scaling using both toroidal 
(squares) and spherical (circles) simulation topologies. Plotted are the 
fugacities at which $\chi$ and $Y_8^\pm$ attain their extrema versus $1/l$, 
with $l=\sqrt{V}$ the lateral extension of the system; the lines are linear 
fits. For systems of finite size, the results differ significantly between the 
different topologies, but agree on the value of $\zcr$ in the thermodynamic 
limit.}
\label{fig:fsscompare}
\end{center}
\end{figure}

In the thermodynamic limit, phase transition properties should not depend on the 
simulation topology. So as a test for the validity of using a spherical 
topology, rather than the more common toroidal one, we compare predictions for 
the critical point. Simulating at fugacities around the critical point, and 
using histogram reweighting \cite{ferrenberg.swendsen:1988}, we verified that in 
both cases the same value of the critical fugacity $\zcr$ is obtained, as well 
as critical exponents consistent with 2D Ising universality. 
\fig{fig:fsscompare} shows the result of a finite-size scaling study along the 
lines of \olcite{orkoulas.fisher.ea:2001}. For both topologies, we have plotted 
the fugacities at which the susceptibility $\chi$ attains its maximum versus 
$1/l$. Here, $l=\sqrt{V}$ denotes the \ahum{length} of the system. Also shown 
are the fugacities at which the generalized susceptibility $Y_8^\pm$ attains its 
global extrema, with $Y_8^\pm$ defined in \olcite{orkoulas.fisher.ea:2001}. The 
lines are linear fits, which capture the data well, consistent with the exact 2D 
Ising value $\nu=1$ of the correlation length critical exponent. Clearly, in 
finite systems, the results between the two topologies differ, but both 
extrapolate to a common value $\zcr=1.717(2)$ in the thermodynamic limit (the 
error reflects the scatter between individual scaling results). Furthermore, in 
both topologies, we find that the maximum value of the susceptibility increases 
with the length of the system as $\chi_{\rm max} \propto l^{\gamma/\nu}$, with 
$\gamma$ the critical exponent of the susceptibility. We obtain $\gamma_t 
\approx 1.754$ and $\gamma_s \approx 1.743$, for the toroidal and spherical 
topology, respectively, which both compare well to the exact 2D Ising value 
$\gamma = 7/4$. Our estimate of $\zcr$ for the 2D WR mixture is close to the one 
reported in \olcite{johnson.gould.ea:1997}, but upon careful inspection does 
underestimate it (by about 0.5\%). Interestingly, one of us (RV) has experienced 
a similar disagreement for the 3D WR mixture as well \cite{vink:2006}. The 
origin of the discrepancy is not clear.

\subsection{The fate of exponential slowing down}

\begin{figure}
\begin{center}
\includegraphics[width=\figwidth, viewport=55 75 750 550]{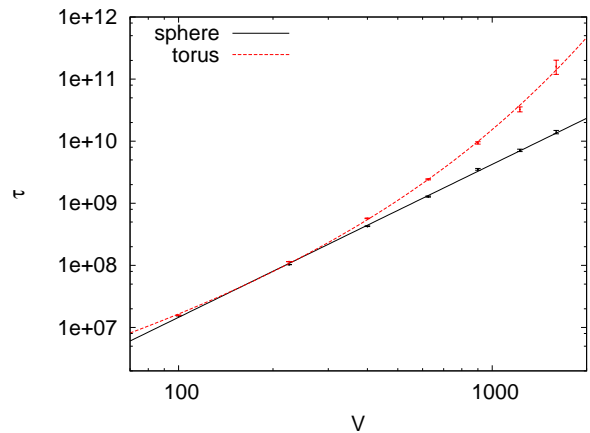}
\caption{Main result of this paper: shown is the number $\tau$ of MC steps 
needed to traverse from $\rho_A=0$ to $\rho_A=3.5$ and back as a function of the 
system area $V$ in biased simulations using toroidal (dashed line) and spherical 
(solid line) boundary conditions. Note the double-logarithmic scale! The 
important result to take from this figure is that $\tau$ increases strongly with 
$V$ (presumably exponentially) on the torus, but only weakly 
(power law) on the sphere. The data were obtained for the 2D WR 
mixture at fugacity $z=2.5$; the exponent of the power law for the spherical 
topology equals $\alpha \approx 2.5$, which is remarkably close to $\alpha_{\rm 
RW}=2$ of a true random walk.}
\label{fig:corrtimes}
\end{center}
\end{figure}

We now come to the main result of this paper, where we consider exponential 
slowing down at a first-order phase transition. Our hope is that, by using a 
spherical topology, exponential slowing down can be largely eliminated. To this 
end, we set the fugacity to $z=2.5$ which is well above $\zcr$, and so the 
transition is strongly first-order. We remind the reader that our simulations 
use a bias function $w(N_A)$ to overcome the free energy barrier of interface 
formation, and also that $w(N_A)$ is {\it a priori} unknown. Hence, for a number 
of system sizes $V$, accurate bias functions $w(N_A)$ were first obtained using 
successive umbrella sampling \cite{virnau.muller:2004}, for both toroidal and 
spherical topologies. Next, biased simulations were performed using the (now 
known) bias functions, and the number of MC steps $\tau$ needed to traverse from 
$\rho_A=0$ to $\rho_A=3.5$ and back was measured; the reader can verify in 
\fig{fig:probdist} that this range is sufficient to sample both the vapor and 
liquid peaks.

The resulting $\tau$ data are collected in \fig{fig:corrtimes}. The most 
striking feature of the plot is that for a toroidal topology $\tau$ indeed 
increases faster than a power law with $V$. Following the discussion in the 
Introduction, we attribute this slow down to free energy barriers associated 
with the droplet-strip shape transition, which are not overcome by the bias 
function $w(N_A)$. The data for the spherical topology, in contrast, do not 
reveal any exponential slow down for the system sizes considered here but rather 
a power law increase; approximately $\tau \sim V^\alpha$, with $\alpha \approx 
2.5$. This is still a slow down compared to a perfect random walk, for which 
$\alpha_{\rm RW}=2$, but not an exponential one. Possibly, the remaining slow 
down is due to nucleation events. Comparing the values of $\tau$ between the two 
topologies, it is striking that already for system area $V=1600$, $\tau$ on a 
torus is ten times that of $\tau$ on a sphere.

\subsection{Wang-Landau sampling}

\begin{figure}
\begin{center}
\includegraphics[width=\figwidth, viewport=55 75 750 550]{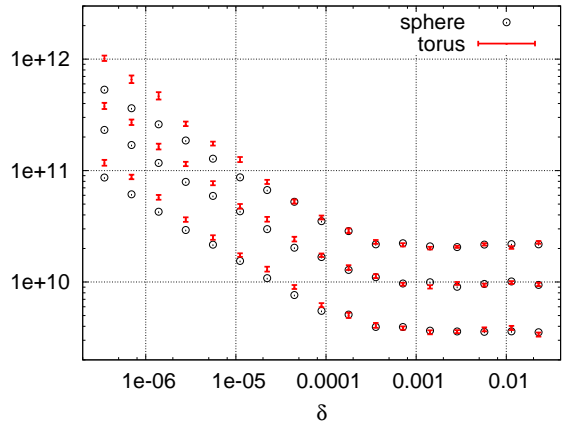}
\caption{Number of MC moves required to complete one WL iteration, using a 
toroidal (vertical bars) and spherical (circles) simulation topology, on double 
logarithmic scales. Results are shown for system areas $V=900, 1600, 2500$ 
(bottom to top). Note that during WL sampling the larger values of $\delta$ are 
sampled first. The data were obtained for the 2D WR mixture at fugacity 
$z=2.5$.}
\label{fig:WLtimes}
\end{center}
\end{figure}

\begin{figure}
\begin{center}
\includegraphics[width=\figwidth]{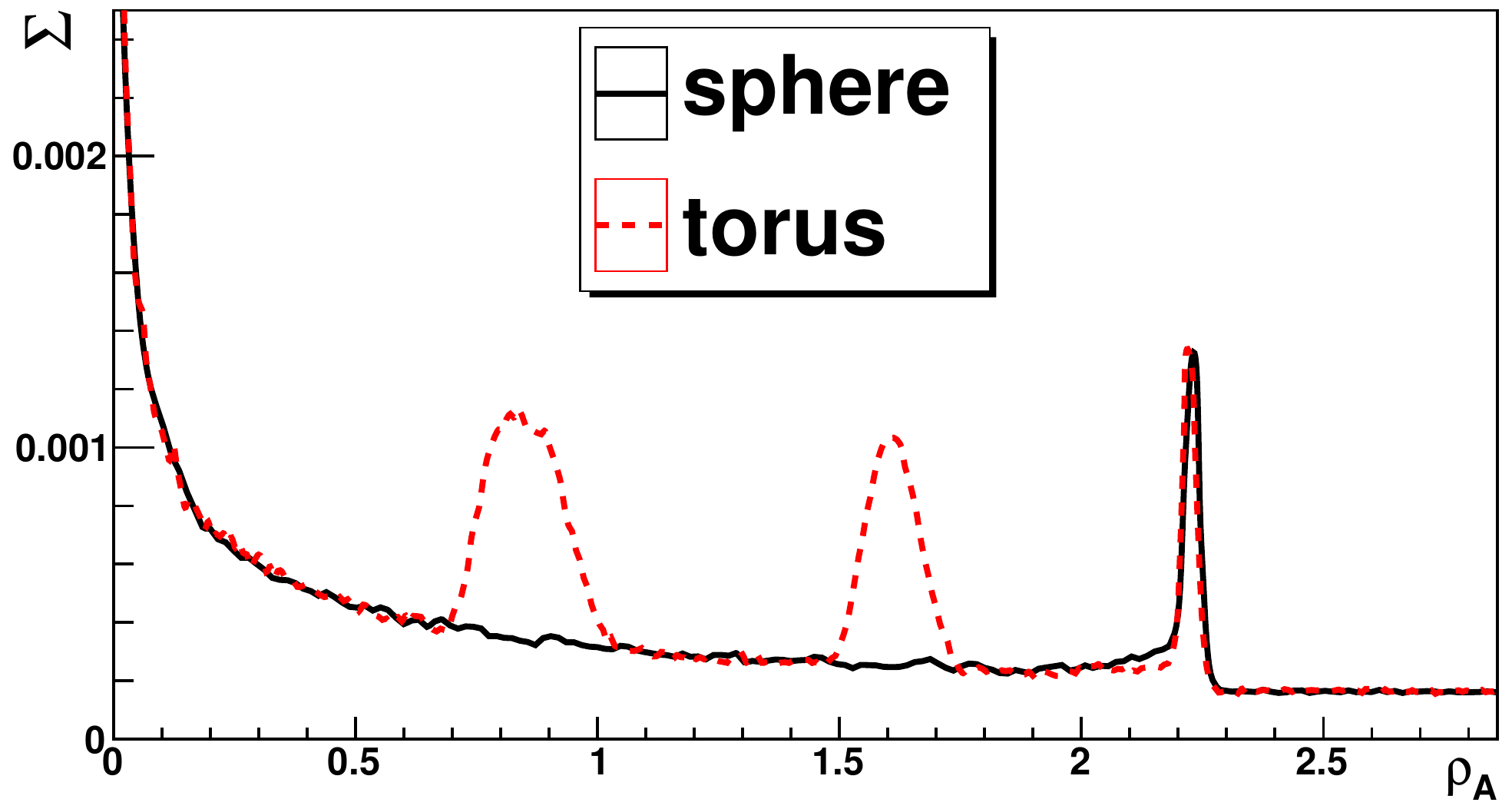}
\end{center}
\caption{Variation of the statistical error $\Sigma$ in $\Delta w(N_A)$ versus 
$\rho_A$, as obtained in WL simulations on toroidal (dashed line) and spherical 
(solid line) topologies. For clarity, intervals of 20 error estimates are 
combined to a single average. The error bars on the toroidal data represent 
jackknife errors for the average of this re-binning; error bars for the 
spherical topology are omitted. The data were obtained for the 2D WR mixture at 
fugacity $z=2.5$ and system area $V=2500$.}
\label{fig:WLerrorcompare}
\end{figure}

\begin{table}
\begin{tabular}{c|ccccc}
\textbf{V} & $\ \ \left< \rho_A \right> \ $ & $\ m_1 \ $ & $\ m_2\ $ & $\ m_3\ $ & $\ m_4\ $ \\
\hline
900  & $1.42$ & $1.69$ & $1.69$ & $3.44$ & $1.21$ \\
1600 & $2.92$ & $3.83$ & $3.83$ & $6.51$ & $3.66$ \\
2500 & $2.67$ & $3.32$ & $3.32$ & $4.70$ & $7.80$ 
\end{tabular}
\caption{Relative accuracy of selected physical observables obtained using WL 
sampling on spherical and toroidal topologies for several system areas $V$. The 
ratios all exceed unity, implying that the spherical topology is to be 
preferred. The data were obtained for the 2D WR mixture at fugacity $z=2.5$.}
\label{tab:WLmoments}
\end{table}

Having shown that the tunneling time $\tau$ using a spherical topology can be 
much smaller compared to that on a torus, a natural next question is whether 
this improvement also increases the performance of the algorithms used to 
construct $w(N_A)$. One such algorithm is Wang-Landau (WL) sampling 
\cite{wang.landau:2001}. Here, the bias function is initially set to $w(N_A)=0$, 
and one proceeds to simulate as explained in Section~\ref{sec:bias}. Each time a 
state with $N_A$ particles of type A is visited, the corresponding value of the 
bias function is decreased by a modification term: $W(N_A) \to W(N_A) - \delta$. 
This reflects the idea that the current number of $A$ particles is just being 
found a bit more probable than previously expected and hence needs a smaller 
bias. One continues to simulate until all particle numbers $N_A$ over the range 
of interest have been visited sufficiently often \cite{flat}, which completes 
one WL iteration. At this point, the modification term is reduced, say, $\delta 
\to \delta/2$, and the next iteration is started. By using a large modification 
term at first, say $\delta=1$, one ensures that all states will be visited in 
relatively short times. At later stages, when $\delta$ is small, changes to the 
bias function become negligible, and the algorithm is said to have converged. 
Clearly, for the performance of this algorithm, it helps if the simulation 
traverses the density range of interest as quickly as possible. This is 
obviously related to the tunneling time $\tau$, which is significantly reduced 
on a spherical topology, and so we expect an increased performance for WL 
sampling too.

To test the performance of WL sampling on toroidal and spherical topologies, the 
bias function $w(N_A)$ was measured independently $40$ times using system sizes 
$V=900, 1600, 2500$ (by independent we mean that each WL run was started with 
its own unique stream of random numbers). In \fig{fig:WLtimes}, the average 
number of MC steps needed to complete one WL iteration is plotted as a function 
of the modification term $\delta$. In the early stages, where $\delta$ is large, 
the number of MC steps is effectively identical. In this regime $\delta$ is so 
large that the simulations on the torus are still easily pushed through the 
regime of the droplet-strip transition, and hence there is no noticeable 
difference. However, at later stages, where $\delta$ is small, the number of 
required MC steps is significantly less on the sphere, as expected, since here 
the sphere simulations benefit from the improved diffusion behavior demonstrated 
in \fig{fig:corrtimes}.

Hence, late-stage WL iterations indeed complete faster on a spherical topology, 
compared to a toroidal one. Next, it remains to be shown that actual physical 
observables are also more accurately obtained. To this end, we consider the 
average density of A particles $\avg{\rho_A}$, and the first four central 
moments
\beq
 m_n = \left< \right. 
 \left| \rho_A - \left< \rho_A \right> \right|^n \left. \right>.
\eeq
Note that $\avg{\rho_A}$ and $m_n$ are trivially obtained from the order 
parameter distribution $P_V(N_A|z)$, which in turn is related to the bias 
function $w(N_A) = - \log P_V(N_A|z)$. For each topology, we thus have a set of 
$40$ estimates per observable. Using the jackknife method 
\cite{newman.barkema:1999}, one can derive the statistical error $\Sigma$ in 
each observable, and the \ahum{best} topology is the one with the smallest 
$\Sigma$. In Table~\ref{tab:WLmoments}, the ratio of errors $\Sigma_t / 
\Sigma_s$ is shown for each observable, with $\Sigma_t$ ($\Sigma_s$) the error 
as obtained on the torus (sphere). The point to take from the table is that the 
ratios all exceed unity, meaning that the data from the spherical topology are 
more reliable. In combination with the findings of \fig{fig:WLtimes}, we 
conclude that WL simulations on spherical topologies are overall more efficient.

Finally, we demonstrate that the enhanced performance of WL sampling on 
spherical topologies can indeed be attributed to the absence of the 
droplet-strip shape transition. To this end, we consider the difference
\beq
 \Delta w(N_A) \equiv w(N_A + 1) - w(N_A)
\eeq
between adjacent weights. Again using the jackknife method, we estimated the 
statistical error $\Sigma$ in $\Delta w(N_A)$ from the set of $40$ simulations 
for each topology. Plotted in \fig{fig:WLerrorcompare} is $\Sigma$ versus 
$\rho_A$. The striking feature is that on the torus, $\Sigma$ displays two 
extra large peaks, which are completely absent on the sphere. These extra 
peaks in the torus data reflect the sampling difficulties arising from the 
droplet-strip transition. Away from the droplet-strip transitions, both 
topologies yield essentially the same statistical error, as expected. Note also 
the excellent agreement between both topologies regarding the peak on the far 
right of the graph: this peak reflects a sampling problem arising from 
nucleation, which indeed should occur in both topologies. In principle, a 
nucleation peak should also be visible on the far left of the graph, but due to 
the high fugacity used, this peak is probably squeezed onto the vertical axes.

\subsection{Successive umbrella sampling}

\begin{table}
\begin{tabular}{c|ccccc}
\textbf{V} & $\ \ \left< \rho_A \right> \ $ & $\ m_1 \ $ & $\ m_2\ $ & $\ m_3\ $ & $\ m_4\ $ \\
\hline
 900  & $1.61$ & $2.61$ & $2.62$ & $6.8$  & $1.83$ \\
 1600 & $3.47$ & $6.34$ & $6.34$ & $14.6$ & $2.67$ \\
 2500 & $4.43$ & $11$  & $11.1$  & $30.1$ & $11.2$
\end{tabular}
\caption{The analogue of Table~\ref{tab:WLmoments} but this time using 
successive umbrella sampling \cite{virnau.muller:2004}.}
\label{tab:SUSmoments}
\end{table}

Another algorithm that can be used to construct the bias function $w(N_A)$ is 
successive umbrella sampling (SUS) \cite{virnau.muller:2004}. When SUS is 
implemented on a spherical topology we also observe an increase in performance. 
In Table~\ref{tab:SUSmoments}, we display some typical benchmarks regarding the 
accuracy of a number of physical observables. Compared to WL simulations, the 
performance increase in SUS simulations appears to be even more pronounced. A 
possible explanation may be that, unlike in WL sampling, each state $N_A$ was 
simulated using the same number of MC steps for both geometries.

\section{Conclusions}

In conclusion, we have shown that the droplet-strip transition, which is 
inherent to toroidal systems undergoing a first-order phase transition, acts as 
barrier making simulations of large systems increasingly harder. For the 2D WR 
mixture investigated here, the droplet-strip transition can be easily, and at 
the cost of a constant fraction of CPU time, be eliminated by simulating the 
system on the surface of a sphere. We have shown that for WL sampling 
\cite{wang.landau:2001} and SUS \cite{virnau.muller:2004}, simulations on the 
surface of a sphere yield better results. The droplet-strip transition is a 
general feature, also in 3D, appearing at all first-order phase transitions 
studied on toroidal topologies. Hence, the use of a spherical topology is 
expected to be beneficial in a great number of other systems also.

The 2D implementation sketched here should quite straightforwardly extend to 3D 
\cite{citeulike:6008340}, and to other (short-ranged) interactions also; obvious 
candidates are the hard-core square-well fluid, the (cut-and-shifted) 
Lennard-Jones fluid, and colloid-polymer mixtures. However, models in which the 
pair potential is a more complicated function of the distance may require the 
use of computationally expensive trigonometric functions, which were 
successfully circumvented in the present implementation. Even so, this 
additional computational effort should be outbalanced by the elimination of 
exponential slowing down, provided the systems are large enough.

Some models may not be so easily transferable to the sphere, in particular when 
particle orientation comes into play. An example is the 2D Zwanzig model 
\cite{zwanzig:1714} of horizontally or vertically aligned hard rods. While on 
the torus one can uniquely speak of horizontal and vertical directions, this is 
prevented by the intrinsic curvature on the sphere. Note also that the advantage 
of using a spherical topology is to eliminate exponential slowing down at {\it 
first-order} transitions. Around the critical point, where the transition is 
{\it continuous}, we did not see much advantage using the spherical topology.

A different application where the use of a spherical topology may be beneficial 
is in the simulation of droplets \cite{citeulike:4503179, 
virnau.macdowell.ea:2003, citeulike:5076926}. One problem of using a toroidal 
topology is that the maximum size of the droplet that one can simulate is 
limited to the point where the droplet-strip transition takes place 
\cite{citeulike:5076926}. On a spherical topology, this problem is circumvented. 
Finally, we would like to point out that phase separation on the surface of a 
sphere is also realized experimentally in giant vesicles 
\cite{citeulike:6008542}; confocal microscopy images of the latter qualitatively 
resemble the simulation snapshots of \fig{fig:sphere_sysview}.

\subsection*{Acknowledgments}

This work was supported by the {\it Deutsche Forschungsgemeinschaft} under the
Emmy Noether program (VI~483/1-1).

\bibliography{mc1975,footnotes}

\begin{thebibliography}{10}

\bibitem{aarts.schmidt.ea:2004}
D.~G. Aarts, M.~Schmidt, and H.~N. Lekkerkerker, Science \textbf{304}, 847
  (2004).

\bibitem{poon:2004}
W.~C.~K. Poon, Science \textbf{304}, 830 (2004).

\bibitem{fsslit}
The literature on finite-size scaling is too extensive to be cited here; for a
  recent review on the use of such methods in soft matter systems see:
  R.L.C.~Vink, Soft Matter, 2009, DOI: 10.1039/b912135h.

\bibitem{berg.neuhaus:1992}
B.~A. Berg and T.~Neuhaus, Physical Review Letters \textbf{68}, 9+ (1992).

\bibitem{widom.rowlinson:1970}
B.~Widom and J.~S. Rowlinson, The Journal of Chemical Physics \textbf{52}, 1670
  (1970).

\bibitem{citeulike:3850776}
A.~R. Honerkamp-Smith, P.~Cicuta, M.~D. Collins, S.~L. Veatch, M.~den Nijs,
  M.~Schick, and S.~L. Keller, Biophysical Journal \textbf{95}, 236+ (2008).

\bibitem{citeulike:5076327}
K.~Leung and R.~K.~P. Zia, Journal of Physics A: Mathematical and General
  \textbf{23}, 4593 (1990).

\bibitem{citeulike:4503186}
T.~Neuhaus and J.~S. Hager, Journal of Statistical Physics \textbf{113}, 47
  (2003).

\bibitem{citeulike:2414151}
G.~Torrie and J.~Valleau, Journal of Computational Physics \textbf{23}, 187
  (1977).

\bibitem{virnau.muller:2004}
P.~Virnau and M.~Muller, The Journal of Chemical Physics \textbf{120}, 10925
  (2004).

\bibitem{wang.landau:2001}
F.~Wang and D.~P. Landau, Physical Review Letters \textbf{86}, 2050+ (2001).

\bibitem{vink.schilling:2005}
R.~L.~C. Vink and T.~Schilling, Phys. Rev. E \textbf{71}, 051716 (2005).

\bibitem{allen.tildesley:1989}
M.~P. Allen and D.~J. Tildesley, \emph{Computer Simulation of Liquids} (Oxford
  University Press, England, 1989).

\bibitem{binder:1982}
K.~Binder, Phys. Rev. A \textbf{25}, 1699 (1982).

\bibitem{berg.hansmann.ea:1993}
B.~A. Berg, U.~Hansmann, and T.~Neuhaus, Phys. Rev. B \textbf{47}, 497 (1993).

\bibitem{citeulike:4609411}
J.~S. Rowlinson, Journal of Physics A: Mathematical and General \textbf{17},
  L357 (1984).

\bibitem{ferrenberg.swendsen:1988}
A.~M. Ferrenberg and R.~H. Swendsen, Phys. Rev. Lett. \textbf{61}, 2635 (1988).

\bibitem{orkoulas.fisher.ea:2001}
G.~Orkoulas, M.~E. Fisher, and A.~Z. Panagiotopoulos, Physical Review E
  \textbf{63}, 051507+ (2001).

\bibitem{johnson.gould.ea:1997}
G.~Johnson, H.~Gould, J.~Machta, and L.~K. Chayes, Physical Review Letters
  \textbf{79}, 2612 (1997).

\bibitem{vink:2006}
R.~L.~C. Vink, The Journal of Chemical Physics \textbf{124}, 094502+ (2006).

\bibitem{flat}
In practice this is determined using a \ahum{flatness} criterion, see for
  example \olcite{citeulike:278331}.

\bibitem{newman.barkema:1999}
M.~E.~J. Newman and G.~T. Barkema, \emph{Monte Carlo Methods in Statistical
  Physics} (Clarendon Press, Oxford, 1999).

\bibitem{citeulike:6008340}
J.~M. Caillol, The Journal of Chemical Physics \textbf{99}, 8953 (1993).

\bibitem{zwanzig:1714}
R.~Zwanzig, The Journal of Chemical Physics \textbf{39}, 1714 (1963).

\bibitem{citeulike:4503179}
K.~Binder, Physica A: Statistical Mechanics and its Applications \textbf{319},
  99 (2003).

\bibitem{virnau.macdowell.ea:2003}
L.~G. Macdowell, P.~Virnau, M.~M\"{u}ller, and K.~Binder, The Journal of
  Chemical Physics \textbf{120}, 5293 (2004).

\bibitem{citeulike:5076926}
M.~Schrader, P.~Virnau, and K.~Binder, Physical Review E (Statistical,
  Nonlinear, and Soft Matter Physics) \textbf{79}, 061104+ (2009).

\bibitem{citeulike:6008542}
H.~M. Seeger, M.~Fidorra, and T.~Heimburg, Macromolecular Symposia
  \textbf{219}, 85 (2005).

\bibitem{citeulike:278331}
F.~Wang and D.~P. Landau, Physical Review E \textbf{64}, 056101+ (2001).

\end{thebibliography}

\end{document}